\documentclass[aps,showpacs,twocolumn,floats,psfig,superscriptaddress]{revtex4-1}

\usepackage{amsmath,amssymb,amsfonts,amsthm,bbm} 
\usepackage{dcolumn} 
\usepackage{graphicx}
\usepackage{epstopdf}
\usepackage{subfigure}
\usepackage{tikz}
\usepackage{color}
\usepackage{verbatim}
\usepackage{layout} 
\usepackage{mathrsfs}  
\usepackage{import}
\usepackage{psfrag}
\usepackage{pstricks,pst-node}
\usepackage[margin=0cm,font=small]{caption}

\newcommand{\C}{\mathbb{C}}

\newcommand{\Z}{\mathbb{Z}}

\newcommand{\caN}{{\mathcal N}}

\newcommand{\str}{ |}
\newcommand{\e}{ \mathrm{e}}
\newcommand{\ed}{\mathrm{e}}

\newcommand{\norm}{ \|}
\newcommand{\beq}{ \begin{equation} }
\newcommand{\eeq}{ \end{equation} }
\newcommand{\bea}{ \begin{eqnarray} }
\newcommand{\eea}{ \end{eqnarray} }
\newcommand{\nn}{ \nonumber }

\newcommand{\ep}{ \epsilon }

\begin{document}

%TITLE
\title{Absence of many-body mobility edges}
\author{Wojciech De Roeck}
\affiliation{Instituut voor Theoretische Fysica, KU Leuven, Celestijnenlaan 200D, B-3001 Leuven, Belgium}
\affiliation{Kavli Institute for Theoretical Physics, Kohn Hall, University of California, Santa Barbara, California 93106-4030, USA}
\author{Francois Huveneers}
\affiliation{Kavli Institute for Theoretical Physics, Kohn Hall, University of California, Santa Barbara, California 93106-4030, USA}
\affiliation{CEREMADE, Universit\'e Paris-Dauphine, 75775 Paris Cedex 16, Paris, France}
\author{Markus M\"uller}
\affiliation{Kavli Institute for Theoretical Physics, Kohn Hall, University of California, Santa Barbara, California 93106-4030, USA}
\affiliation{Paul Scherrer Institut, CH-5232 Villigen PSI , Switzerland}
\affiliation{The Abdus Salam International Center for Theoretical Physics, Strada Costiera 11, 34151 Trieste, Italy}
\affiliation{Department of Physics, University of Basel, Klingelbergstrasse 82, CH-4056 Basel, Switzerland}
\author{Mauro Schiulaz}
\affiliation{SISSA  and INFN, Via Bonomea 265, 34136 Trieste, Italy}

\date{\today}
\doi{10.1103/PhysRevB.93.014203}

\begin{abstract}
\noindent
Localization transitions as a function of  temperature require a many-body mobility edge in energy, separating localized from ergodic states. We argue that this scenario is inconsistent because local fluctuations into the ergodic phase within the supposedly localized phase can serve as mobile bubbles that induce global delocalization. Such fluctuations inevitably appear {with a low but finite density anywhere} in any typical state. We conclude that the only possibility for many-body localization to occur are lattice models that are localized at all energies. 
Building on a close analogy with a {model of assisted two-particle hopping}, where interactions induce delocalization, we argue why hot bubbles are mobile and do not localize upon diluting their energy. 
{Numerical tests of our scenario show that previously reported mobility edges  cannot be distinguished from finite-size effects.}
\end{abstract}

\maketitle

% SECTION
\section{Introduction}
It is now almost mathematically  proven that many-body localization, i.e., the absence of long-range transport in a thermodynamic many-body system, occurs in certain one-dimensional quantum lattice models at any energy density if sufficiently strong quenched disorder is present~\cite{Imbrie}. In this case, many-body localization (MBL) comes along with a complete set of conserved quasi-local quantities~\cite{integrals1,integrals2,muellerros}. However, it remains less clear whether the originally predicted localization transition at finite temperature~\cite{BaskoAleinerAltshuler,Mirlin} exists as a genuine dynamical phase transition defining a sharp many-body mobility edge in energy density. Even though several numerical investigations in small one-dimensional (1D) systems have reported such mobility edges~\cite{Pollman,Alet,Laumann}, studies in larger systems did not find similar evidence~\cite{Carleo, Rigol} {and, moreover, linked-cluster analysis \cite{devakul} of the numerical data hint that the extent of the localized phase has been vastly overestimated.} Furthermore, recent theoretical considerations~\cite{bubbles, bubbles2, muellerros, SchiulazMueller, discussionhuserahul}  have raised doubts about non-perturbative effects which might reduce the putative transition to a crossover. 
A related open issue concerns the many-body analog of Mott's argument, which forbids the coexistence  of localized and delocalized states at the same energy in single-particle problems. 

In this paper, we address these issues, which are  fundamental  for a complete understanding of localization, equilibration, and transport in closed many-body quantum systems. 
We argue that for systems with short-range interactions, many-body mobility edges cannot exist, thus ruling out sharp transitions from a conducting to a completely insulating phase as a function of temperature. 
These considerations also imply a strong many-body version of Mott's argument, which rules out the coexistence of localized and delocalized states, even at extensively different energies. A simple generalization of our considerations rules out mobility edges as a function of any extensive thermodynamic parameter, such as particle number or magnetization.

Our paper is organized as follows. 
We introduce all our arguments at a non-technical level in Sec. \ref{sec: summary}, and argue that local hot thermal spots, dubbed \emph{bubbles}, constitute a mechanism for global delocalization. 
Section \ref{sec: formal bubbles} contains a more detailed presentation of the argument, while Sec. \ref{sec: caveats} is devoted to the analysis and discussion of potential caveats.
Our numerical results are presented in Sec. \ref{sec: numerics}: 
By a careful analysis of the model considered by Kj\"all {\em et al.}\cite{Pollman}, we show that currently available system sizes are too small to host a truly thermal bubble, and, hence, that existing numerical data do not contradict our theory. 
To conclude, in Sec. \ref{sec: conclusion}, we discuss the physical consequences of our analysis, and point out, in which physical systems it allows a genuine MBL phase to exist.
More detailed information and discussions are relegated to three appendixes: Appendix~\ref{sec: assisted hopping} presents  numerical results demonstrating delocalization via rare events in a two-particle model. In Appendixes~\ref{sec: quenched spots} and \ref{sec: single particle}, we discuss why the idea of bubbles as rare fluctuations of energy or density neither apply to many-body systems with disorder  which localizes the full spectrum, nor to single-particle problems.

% SECTION
\section{Summary of the arguments}\label{sec: summary}
In this section, we first discuss a simple {model of assisted two-particle hopping}, which illustrates several important features that this problem has in common with the rare events that induce delocalization in many-body systems
 (see Appendix \ref{sec: assisted hopping} for numerical results).  For this model, we show how rare local fluctuations induce hybridization among putatively localized states.
Then, we turn to our main topic, general many-body systems. We explain in non-technical terms how those rare events wash out mobility edges whenever there is an ergodic state at some finite energy or particle density (or, more generally, if there are ergodic states in any region of the parameter space for the extensive thermodynamic quantities).
Finally, we discuss some potential caveats, and  argue why they are benign.

\subsection{Assisted hopping model}\label{subsec: assisted hopping summary} 
\begin{figure}[h!]
\begin{center}
\includegraphics[scale=0.8]{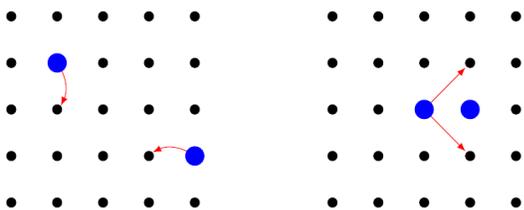}
\end{center}
\caption{
\label{fig: assisted hopping}
Hopping processes allowed by the Hamiltonian of Eq.~(\ref{H}) in $d=2$. 
Left panel: single-particle hopping. 
Right panel: assisted hopping for the left particle due to the presence of the right particle. 
}
\end{figure}

Consider particles in a hypercubic lattice of linear size $L$, hopping with amplitude $t_1$ between nearest-neighbor sites, and subject to a disorder potential $\epsilon_x$, i.i.d. uniformly in $[-W,W]$. 
A particle on site $x$ interacts with others by inducing an assisted hopping of strength $t_2$ along the diagonals of plaquettes that $x$ belongs to
\bea
\label{H}
H&=& - t_1\sum_{\langle x,y\rangle} (c^\dagger_x c_y+ {\rm H.c.})+\sum_x \epsilon_x n_x\\
&& - t_2 \sum_{x}\sum_{s,s'=\pm 1} \sum_{1\leq \alpha<\beta\leq d} n_x(c^\dagger_{x+s\vec{e}_\alpha} c_{x+s'\vec{e}_\beta}+ {\rm H.c.}), \nn
\eea
where $\vec{e}_{1,\cdots,d}$ are lattice unit vectors. This may describe the effect of a lattice distortion brought about by the presence of the first particle. This model is illustrated in Fig.~\ref{fig: assisted hopping}. 

Let us first focus on the particular case where there are only two particles in the system. 
We consider parameters $t_1\ll W$, for which the single-particle problem is localized in the whole spectrum. 
For $t_2\gg W$, the two-particle problem has several interesting features: In dimensions $d>2$, the assisted hopping term induces a delocalization of close pairs {which will move together diffusively as a composite light particle and overcome Anderson localization. This effect is related to the interaction-induced increase of the localization length in sufficiently weakly localized systems~\cite{Shepelyansky,Imry}. A single-particle analog of the phenomenon is the solvable case of two coupled Bethe lattices~\cite{XieMueller}}. 
The delocalization in (\ref{H}) seems natural, since all configurations of two particles at distance one are strongly resonant with each other. 
They thus form a percolating, delocalized resonant subgraph in configuration space, which supports delocalized wave functions with inverse participation ratios that vanish as the inverse volume. 
We have numerically confirmed this delocalization effect [see  Appendix~\ref{sec: assisted hopping} (see also Refs.~\cite{Shepelyansky3d,Jacquod} for similar observations)].   
In a system of only two particles, the eigenstates come in two kinds: the overwhelming number of states is strongly concentrated on a configuration with two distant, immobile particles. 
Only a vanishing fraction of order $\left[\log(L)\right]^d/L$ of all two-particle eigenstates are delocalized as dynamically bound, mobile pairs. As we will discuss in more detail in Sec.~\ref{sec: formal bubbles}, it would be misleading to think that two particles that  start off together will always lose sight of each other after some time and localize far from each other. Instead, we will argue that the delocalization channel via pair configurations, and thus delocalized eigenfunctions, are robust.

In this two-particle model, localized and delocalized states coexist at the same energy. In contrast to generic single particle problems, this is possible here because the matrix elements that couple the two kinds of states through a random perturbation of the Hamiltonian are typically exponentially small in the system size. They are thus negligible as compared to the relevant level spacings and hence do not  hybridize the two types of states.

Let us now discuss how a finite density of particles modifies the situation. In the thermodynamic limit, there is a finite density of close pairs in typical configurations. These pairs diffuse through the sample. Initially well-isolated and localized particles scatter inelastically off these pairs and thus move as well, leading to complete delocalization. Even in exponentially rare configurations where initially all particles are far from each other, particles eventually tunnel together and decay into the continuum of diffusive pair states. We thus do not expect any localized eigenstates to survive at finite density.

\subsection{Many-body systems:\\ Delocalization from rare bubbles}
The argument of Basko {\em et al.}~\cite{BaskoAleinerAltshuler} for a localization transition as a function of temperature, i.e., a many-body mobility edge, builds on the idea that conduction can set in only if the energy density exceeds a critical level essentially everywhere in the sample. However, this neglects the {fact that local fluctuations away from the average energy density generally cause a breakdown of perturbation theory and may induce delocalization. Indeed, the perturbative analysis of Ref.~\cite{BaskoAleinerAltshuler} focuses on scattering processes where at every vertex an additional particle-hole pair is created, which was justified by the parametrically larger number ($K$) of such diagrams as compared to diagrams that preserve or even reduce the number of particle-hole excitations. However, going beyond this approximation and starting from a generic initial low energy state, after a {\em finite} number of steps in the perturbative expansion one couples to configurations which contain hot, internally ergodic bubbles, for which perturbation theory is not controllable anymore. Our arguments following suggest that the further perturbation theory on this diagrammatic branch diverges and cannot be resummed. We conjecture that a parametrically small (in $K$) but finite conductivity results in that case.}

{In physical terms,} we argue that delocalization occurs as soon as finite, but mobile excitations exist, even if they {occur with very low density}. These highly excited fluctuations constitute the analogs of the diffusive pairs in the assisted hopping model discussed above. Examples of such excitations are large, albeit finite regions which are hotter than their environment and thus are internally ergodic. Hereby we assume that interactions are local, so that the internal ergodicity is only a function of the energy contained in that region.  
%\wdr{For the sake of concreteness, we also assume that in our model  the delocalized phase is indeed at higher termperature than the putative localized phase.}

Let us stress already now that the bubble excitations considered in this paper are thermal, and are not tied to {anomalous realizations of the disorder in specific locations}. Therefore, those bubbles can potentially show up in any location in the system. 
We notice that, as far as thermal bubbles are concerned, the strategy of Imbrie~\cite{Imbrie} to show the existence of an MBL phase would fail. 
Indeed, it requires that the location of all possible resonant spots can be determined independently of the state of the system. 
As this is a crucial point, we review in Appendix~\ref{sec: quenched spots}  the treatment of Ref.~\cite{Imbrie} of ergodic spots that are tied to rare disorder realizations in specific locations. In particular we spell out why those do not lead to delocalization of fully MBL systems in $d=1$, while thermal bubbles do.

Let us now assume that at some temperature there is conduction and ergodicity.\cite{foot1}
In typical states and in any given place finite bubbles of sufficiently high temperature (that ensure internal ergodicity) occur with finite probability as spontaneous fluctuations of the energy density. Those are {\em not} tied to a particular local disorder realization. Thus, at any instant of time there exists a possibly very low, but finite density of such ergodic fluctuations. Following, we argue that such excitations are mobile and delocalize the whole system, akin to the diffusing pairs above. From this reasoning it follows that finite conduction at some temperature implies finite conduction at any temperature in thermodynamic systems with local interactions.\cite{foot2}
As a consequence, systems in the continuum should exhibit finite transport at any $T>0$, as they always possess ergodic states at high enough energy (see also the discussion in Ref.~\cite{Rahul}).

To argue for the mobility of bubbles, we show that there exists a resonant, delocalized subset of bubble configurations.
We consider a quantum lattice system with local interactions and a bounded energy density, possessing a putative many-body mobility edge at energy density $\epsilon_c$, such that states below (above) $\epsilon_c$ are localized (ergodic). 
For simplicity, we assume the model to be one-dimensional. 
Now, consider a rare hot bubble of a super-critical energy density at some $\epsilon_2 > \epsilon_c$, surrounded by "cold" regions of energy density $\epsilon_1<\epsilon_c$. If this energy fluctuation is large enough [much larger than a correlation length $\xi(\epsilon_2)$] and decoupled from its surrounding, it is internally ergodic by assumption.

We argue that this state can hybridize with a translate of the bubble by some length $\ell_0> {\rm max}[\xi(\epsilon_1),\xi(\epsilon_2)]$ when the coupling between the hot region and its surrounding is switched on. 
It suffices to show that extending (or shortening) the hot region by a length $\ell_0$ (by heating up or cooling down the neighboring regions of size $\ell_0$) can occur as a resonant transition. The hybridization with the translated bubble then follows from two successive hybridization processes, as illustrated in  Fig.~\ref{fig: Hybridization}. 
For the latter, it is enough to show that changing the energy in the boundary region by a finite amount is a resonant process.
Let $H_1=g O_{h}\otimes O_{c}$ be the interaction term coupling a hot ($h$) and a cold ($c$) region of size $\ell_0$ across their common boundary. Let $\Psi,\Psi'$ be eigenstates in the hot region and $\eta,\eta'$ eigenstates in the cold region. 
For any hot eigenstate $\Psi$ {in a sufficiently large bubble} we can find (many) $\Psi'$ such that
\beq \label{ratio}
\frac{\str \langle \Psi\eta \str H_1\str \Psi'\eta'\rangle\str}{   \str E(\eta)-E(\eta')+E(\Psi)-E(\Psi')\str } \gg 1,
\eeq
because on the one hand, {by the eigenstate thermalization hypothesis (ETH)~\cite{ETH}}, $\str \langle \Psi\ \str O_h\str \Psi'\rangle\str \sim d_h^{-1/2}$ where $d_h$ is the dimension of an appropriate micro-canonical ensemble for the hot bubble at the energy density set by $\Psi$, while the matrix element $\str\langle \eta\str O_{c}\str \eta'\rangle\str =O(1) $ is finite and independent of $d_h$. On the other hand, we can pick $\Psi'$  such that $\str E(\eta)-E(\eta')+E(\Psi)-E(\Psi') \str \leq W/d_h$, where $W$ is the energy width of the ensemble. The ratio in~(\ref{ratio}) thus scales as $\sim d_h^{1/2}$ and grows exponentially with the length of the bubble. It may thus become much larger than unity, indicating a resonant process. This is not surprising:  it merely expresses that a sufficiently large ergodic bubble acts as a bath for small systems coupled to it. It follows that configurations with hot bubbles in different positions hybridize with each other. {We expect that the eigenfunctions of the system hybridize essentially all configurations which are resonantly connected, implying delocalized eigenfunctions. Since it is easy to check that any configuration consistent with global  conservation laws can be reached via resonant processes, we expect that eigenstates also satisfy ETH  in the thermodynamic limit.
While for generic many-body systems our arguments rule out the coexistence of localized and delocalized states, mobility edges are instead well established for one-particle systems. In Appendix~\ref{sec: single particle}, we explain that this does not imply any inconsistency, since our reasonings about bubbles do not apply to one-particle systems.

In Sec.~\ref{sec: formal bubbles}, we will present a more thorough discussion of the properties of the resonant subgraph of configurations. 

\begin{figure}[h!]
\begin{center}
\includegraphics[scale=0.7]{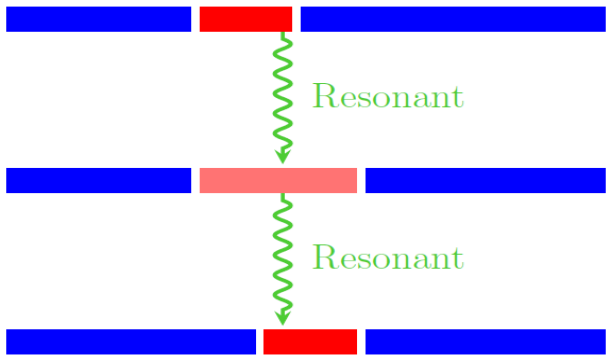}
\end{center}
\caption{
\label{fig: Hybridization}
Hybridization process. 
The state with the bubble on the left (top) hybridizes with the state with the bubble on the right (bottom) via an intermediate state. 
Equation~(\ref{ratio}) shows that the two transitions depicted here are resonant. 
}
\end{figure}

\subsection{Potential caveats: Can bubbles freeze {despite a percolating resonant subgraph?}}\label{subsec: caveats}
We now ask whether processes that have not been taken into account in the previous analysis could impede the hybridization of bubbles. 
Indeed, what was argued up to now can be summarized by saying that there exists a connected subgraph of base states, along which all transitions are resonant. {This subgraph is delocalized in the sense that most member base states differ from each other in degrees of freedom at arbitrary far distances in real space.} These are the base states that contain well-delimited and ergodic bubbles. 
 What happens when we consider the coupling to base states lying off this resonant subgraph?  
However, before we go into this \emph{technical} analysis [see \emph{(b)} below], let us first look at it in an intuitive fashion.
 
 \emph{(a) Complete disappearance of bubbles.}
 If one thinks about the issue in a dynamical setting, rather than as an exercise in spectral perturbation theory for eigenstates, then the following consideration comes up:   Intuitively, hot bubbles should not survive dynamically, but should rather spread, dilute their energy, and eventually localize, so that they could not evolve back to their original hot configuration. If this were true, then surely it would rule out delocalized eigenstates. 
 
However, although such a spreading is indeed entropically favored in real-time dynamics, the above conclusion that hot configurations could therefore not form again is fallacious.   We consider an ensemble of mutually orthogonal initial states of a given energy density. When summing {the projectors onto} these initial states, we recover a thermal (microcanonical) density matrix,  even if the initial states themselves are not thermal,  and hence the mean number of bubbles remains constant in time by the time invariance of the thermal density matrix.  In simple terms: thermal fluctuations do occur, and keep occurring, irrespective of whether the system is localized or not. The only thing that localization could do is to pin the fluctuations at fixed positions. However, our argument is ruling out precisely this option.

\emph{(b) Technical analysis of off-resonant couplings.}
We should convince ourselves that coupling to configurations off the resonant subgraph does not  spoil the resonance of transitions on the graph of bubble configurations. 
Suppose that the spreading mixes the original hot bubble states with states in which the bubble has spread partially, and which thus form a larger local Hilbert space of {\em finite} dimension $d_h' \gg d_h$ \cite{foot3}.
To argue as much as reasonably possible against delocalization,  we suppose that the resonant coupling between hot regions centered in different positions is restricted to the original $d_h$ configurations (because the spread-out bubbles may have lost their ability to translate directly).
 Assuming ergodicity within the larger space of dimension $D_h \equiv d_h+d_h'$, the matrix elements get reduced by a factor $D_h/d_h$ (see Sec. \ref{sec: caveats} for calculations).  However, the minimal denominators decrease by essentially the same factor to $\widetilde W/D_h$, albeit with a slightly larger energy range $\widetilde W$. The ratio $\widetilde W/W$ is bounded, hence it cannot destroy the hybridization if the original bubble was large enough. This contrasts with single-particle problems where sufficiently strong coupling to a bath may induce localization due to a significant increase of the effective bandwidth, as discussed in~\cite{huselocbybath}.

{In summary, the admixture of configurations that are not part of the resonant network cannot prevent the resonant hybridization along the network, but it does increase the timescale for transitions between different positions of the bubble by a factor $D_h/d_h$. This is ultimately very similar as in the assisted hopping model where the possibility of separation of two spatially close particles cannot prevent their finding back together and diffusing further, while it can diminish the pair diffusion constant.}

\emph{(c) Other obstructions to delocalization.}
There are examples of nonergodic behavior that are not straightforwardly captured by the above analysis: $(1)$ single-particle localization in weak disorder in low dimension, where the proliferating amplitude of return to the origin in $d\leq 2$ destroys hybridization at large distances; $(2)$  the "quantum percolation" problem, where we see generation of random self-energies from the structural disorder along barely percolating paths~\cite{Shapir,LoganWolynes};  $(3)$ many-body orthogonality catastrophes, as in spin-boson systems at $T=0$ and related  spin problems at finite $T$~\cite{Stamp}. All these examples rely on specific mechanisms that seem not to be present in our problem. For the case $(1)$, this is addressed explicitly in Sec.~\ref{sec: caveats}. 
Another issue are rare regions with anomalously strong disorder. Those may render transport in $d=1$ subdiffusive \cite{agarwal, Alet2}, but we argue in Sec.~\ref{sec: caveats} that they do not prevent delocalization by bubbles.

\subsection{Analogy with bubbles in kinetically constrained, classical glasses}
It is interesting to draw an analogy between the bubbles discussed here, and bubbles that mediate transport in classical, kinetically constrained models~\cite{BiroliToninelli}. In both cases, mobility is ensured due to rare  fluctuations in local configurations that allow the system to move and explore the phase space ergodically. In generic kinetically constrained models, the diffusion constant remains finite and only vanishes as the system becomes entirely jammed at maximal density. 
Those models are dynamically very similar to disorder-free quantum models or disordered quantum models with a high-temperature ergodic phase, for which our arguments imply the impossibility of genuine and robust many-body localization. 

In special classical glass models, such as the Knight model,~\cite{BiroliToninelli}, certain moves are strictly disallowed. Moreover, the possibility of moves generally depends on a set of configurational constraints, whose number may diverge upon tuning a parameter. In that case, a genuine dynamic glass transition can take place.  
 That transition is vaguely analogous  to the many-body localization transition, tuned as a function of disorder or interaction strength. Indeed, in both cases any finite bubble with whatever optimized properties remains ultimately immobile. 
However, in contrast to the Knight model, which has no quenched disorder, a genuine MBL transition does require quenched disorder. 

Of course, an appropriate quantum version of the Knight model would also be many-body localized, but the corresponding phase would probably not  be robust with respect to  local perturbations of the Hamiltonian since those re-introduce finite amplitudes for moves that were previously exactly suppressed. The latter constitutes an essential ingredient for the sharp glass transition in the kinetically constrained models. Insofar one may consider the kinetically constrained glass phases as fine tuned (relying on certain exactly vanishing transition amplitudes). On the other hand, the robustness of MBL with respect to random local perturbations of the Hamiltonian has  no classical analogy to our knowledge.

% SECTION
\section{Detailed argument for hybridization by bubbles}\label{sec: formal bubbles}
We describe here our argument for hybridization {bubble configurations in eigenstates} at a more formal and detailed level.  
We consider a quantum lattice model with local interactions, having a putative many-body mobility edge. 
For concreteness, we assume that the model is one dimensional and that states below energy density $\epsilon_c$ are (putatively) localized, whereas those above $\epsilon_c$ are ergodic. {We choose the energy density of the bottom of the spectrum as a reference and set it to zero.} We also impose a maximal energy density $\epsilon_m>\epsilon_c$, reflecting the fact that the Hilbert space is locally finite.  

For our argument it is important to have states at disposal that are clearly ergodic or localized in a given finite volume. Therefore, we introduce, somewhat arbitrarily, a window of energy densities $\ep_{c1} <\ep_{c} <\ep_{c2} $ that are too close to critical to be identified clearly as localized or delocalized in nature.  Let $\ell(\ep)$ be the localization length, which diverges as $\ep %\mathop{\rightarrow}\limits_{<} 
\nearrow \ep_c$.  {We express lengths in units of lattice spacings, and energy densities $\ep$ as energy per site.}
We will now coarse grain the model and group $\ell_0$ adjacent sites into "grains". The length $\ell_0$ is chosen such that it is
  {\em (a)} larger than the localization length $\ell(\ep_{c1})$, and {\em (b)} large enough so that the interaction energy between two neighboring grains is small compared to  $\ell_0 \ep_{c1}$, the maximal energy in a localized grain.  The second constraint 
    ensures that the interaction of a low-energy grain ($\ep< \ep_{c1}$) with its surroundings does not trivially suffice to render the grain ergodic, and, similarly, that the interaction of a high-energy grain with its surroundings does not trivially suffice to localize that grain.   This will be satisfied by the choice of a large enough $\ell_0$, since the interactions are local.

The coarse graining provides a useful starting point, from which to proceed with perturbation theory in the inter-grain coupling.
 Within each grain, we compute the eigenstates which come in three kinds: cold (below $\ep_{c_1}$), hot (above $\ep_{c2}$), or intermediate (between $\ep_{c1}$ and $\ep_{c2}$). 
If we consider the Hamiltonian without the interaction between grains, then obviously the eigenstates are products of grain eigenstates. Let us focus on eigenstates at very low energy density $\ep \ll \ep_{c1}$. {Our aim will be to show that even those low-energy states are delocalized.} 
In such states, typically, non-cold grains appear only with a density $\nu$ that tends to $0$ as $\ep/\ep_{c1} \to 0$. Chains of labeled grains such as \ $ccciccchhhiiccc$ serve as "mesostates", with $c/i/h$ standing for cold/intermediate/hot.
We can now write our model as 
\begin{equation}
H= \sum_x \left[ H_0(x)  + H_1({x,x+1})\right],
\end{equation}
where $x$ labels the grains, $H_0(x)$ acts on the Hilbert space $\mathcal{H}(x)$ at grain $x$ only, and $H_1$ {describes} the coupling between neighboring grains. Mesostates are {collections of} eigenstates of the term $H_0$. We now consider switching on the coupling terms and evaluate their effect on the unperturbed eigenstates of $H_0$. 
This procedure is similar to the one followed in Ref.~\cite{Imbrie}. 
First we add the interaction terms between cold grains (see following for what is meant precisely). Since we assumed that $\ell(\ep_{c1}) < \ell_0$, this will not have much effect on the localized eigenstates, which thus remain close to products. Note that by doing this, from the point of view of a typical state at energy density $\ep$, we have {already} added most of the interaction terms. What remains is a small fraction  $\sim 2\nu$ ({which is controlled} by the overall energy density) of {all} interaction terms.
  We now add the interaction terms between hot grains. 
By assumption, sufficiently long stretches of such grains $\ldots hhhh \ldots $ (which we call 'bubbles') are ergodic and we will assume that the resulting hot eigenstates in those bubbles satisfy the eigenstate thermalization hypothesis (ETH). 
The situation at this moment is hence that we have partitioned the Hilbert space into a big direct sum, and the Hamiltonian $H_0$ is block diagonal, with the blocks labeled by mesostates.  Let $P_x^r$ be the projector that restricts the value of $H_0(x)$ so that grain $x$ is of type $r=h,i,c$. With this notation the  interaction terms that have already been added are 
\beq  \label{eq: perturbation term}
  P_{x}^{r'_x} P_{x+1}^{r'_{x+1}} H_1({x,x+1}) P_{x}^{r_x} P_{x+1}^{r_{x+1}},  \eeq
for $(r_x,r_{x+1})=(r'_x,r'_{x+1})=(c,c) $, and for $(r_x,r_{x+1})=(r'_x,r'_{x+1})=(h,h) $. 
Some terms are obviously very small  (because the interaction is local in energy) and seem  irrelevant, namely those corresponding to, $(r_x,r_{x+1})=(c,c), (r'_x,r'_{x+1})=(h,h) $ and with primes and no primes reversed.  The main terms that we will be focusing on are those that allow bubbles to spread and move. Those are terms with
\beq
r_x=r'_x = h, \qquad  \text{and arbitrary $r_{x+1}, r'_{x+1}$},
\eeq
and with $x$ and $x+1$ reversed.
They are at the focus of the next section.

\subsection{Resonant delocalization of bubbles} \label{sec: mobility of bubbles}
Let us now consider states of the following form (hot bubble in a cold environment)
\begin{equation}
cccccccc\underbrace{hh\ldots hh}_{n \, \text{grains}}cccccccc,
\end{equation}
where $n$ is sufficiently large so that the eigenstates in the bubble satisfy ETH. We now argue that this state hybridizes with translates of the bubble when we add some of the missing {coupling} terms: In particular, we want to admix the mesostates (with $x,y$ labeling specific grains)
%\begin{align}
%& \ldots ccchhhhccc  \ldots \,  \leftrightarrow   \,  \ldots  cccchhhhcc  \ldots  \nonumber
%%_{\uparrow H_1}\hspace{-3mm}
%\end{align} 
\begin{align}
& \ldots ccc\underset{x}{h}hhh\underset{y}{c}cc  \ldots \,  \leftrightarrow   \,  \ldots  ccc\underset{x}{c}hhh\underset{y}{h}cc  \ldots
%\nonumber
%_{\uparrow H_1}\hspace{-3mm}
\end{align}
{in which the bubble has been translated by one grain.  More precisely, we mean that most microstates (i.e., eigenstates of the Hamiltonian considered up to now) corresponding to the left mesostate can hybridize with a lot of microstates corresponding to the right mesostate. This in turn strongly suggests} that we should expect all {eigenstates} to delocalize completely over these two mesostates. 
To obtain this, we have {included} the relevant coupling terms~(\ref{eq: perturbation term}) corresponding to two bonds $(x,x+1)$ and $(y-1,y)$.
%\begin{align}
%& ccch_{\str_b}hhh_{\str_a}ccc   \ldots \,  \leftrightarrow   \,  \ldots   ccc_{\str_b}chhh_{\str_a}hcc \nonumber
%%_{\uparrow H_1}\hspace{-3mm}
%\end{align} 
This hybridization process can be broken down into elementary steps, that is, transitions at first order of perturbation theory. First, by energy exchange with the hot region, the cold ($c$) grain at $y$ is heated until it becomes intermediate ($i$) and finally hot ($h$). Second, the $h$ grain at $x$ is cooled down until it becomes $c$, via intermediate stages of $i$. 
Microscopically, let us consider a state $\Phi$ corresponding to the mesostate $ccchhhhccc$ and such that $H_0(y)$ is not far below $\ep_{c1}$.
We will argue that $\Phi$ hybridizes with a lot of states $\Phi'$ corresponding to the mesostate  $ccchhhhicc$ where $r'(y)=i$. If instead $H_0(y)$ is far below $\ep_{c1}$, then it hybridizes with a lot of states $\Psi'$ which still corresponds to $r'(y)=c$ ($ccchhhhccc$), but now with $H_0(y)$ a bit closer to $\ep_{c1}$. (A direct step to an intermediate state might instead require adding too much energy in one transition. Therefore, we split the process into several small heating steps to make sure our argument remains valid also if the interactions are assumed to be strictly local in energy.)
Finally, we need to increase the energy stepwise from $i$ to $h$ at grain $y$. The argument for all these transitions is essentially the same and for the sake of simplicity, we stick to $r(y)=r'(y)=c$. The next subsection shows the flexibility of the argument. 

Obviously, it suffices to take eigenstates in $\Phi,\Phi'$ in the region $[x,y]$  because of the essential product structure (exact at the left edge, approximate at the right edge {around $y$}, because we have already included the coupling between cold regions). They are of the form 
\begin{equation}
\Phi= \Psi \otimes \eta, \qquad \Phi'=\Psi' \otimes \eta',
\end{equation}
where $\eta,\eta'$ are the {unperturbed} eigenstates at grain $y$, while $\Psi,\Psi'$ are  hot bubble states in the region $[x,y-1]$ consisting of $n=y-x$ grains. 
Consider $\Psi'$ such that its energy (evaluated with $H_0$) is within a range $W\sim \ep_m$ of the energy of $\Psi$.  The space spanned by such states has dimension {$d_{h}\approx \exp[s \ell_0 n]$ which grows exponentially in $n$, $s$ being the corresponding entropy density.} Write $H_1(y-1,y)=gO_h \otimes O_{c}$, the first factor acting on $y-1$, the second on $y$.  Assuming ETH, the off-diagonal matrix elements of local operators are given by
\begin{equation}
\str \langle \Psi \str O_h \str \Psi'\rangle\str  \sim 1/\sqrt{d_{h}}.
\end{equation}
%\wdr{I would keep either of the two, but not both, because they are basically the same, I would say. Having the 'and' suggests that we are using more than ETH, which I think is not true}
In other words, the (non-eigenstate) vector $O_h \Psi$ is {essentially a random amplitude superposition} of eigenstates $\Psi'$. 
%The argument goes by connecting the leftmost state via resonant transitions to the rightmost state. It suffices to show that the bubble $hhhh$ can change the state of the $c$ to its right,then it can also change it until the $c$ becomes $H$. Afterwards, it will be clear that in fact, we can connect the rightmost state directly to a bubble translated over a small fraction of its length \wdr{but that is omitted for the time being}.  Let $H_1=g O_{h}\otimes O_{c}$ be the interaction term coupling left ($l$) and right ($r$) across the cut indicated by $\str_a$, let $\Psi,\Psi'$ be eigenstates in the bubble and $\eta,\eta'$ eigenstates in the $c$ right next to it (indeed, in the localized region these eigenstates are essentially products).
Take now $\Delta E:=E(\eta)-E(\eta')$ sufficiently small, i.e.\ not exceeding $W$, then $\str\langle \eta\str O_{c}\str \eta'\rangle\str \sim 1$. In fact, assuring the non-vanishing of $\str\langle \eta\str O_{c}\str \eta'\rangle\str $ is the main reason to choose {$\Delta E$}
sufficiently small.
 We can then find many $\Psi'$  (in fact, $\sim \sqrt{d_{h}}$ of them) such that
\beq \label{eq: ratio}
\frac{\str \langle \Psi\eta \str H_1\str \Psi'\eta'\rangle\str}{   \str \Delta E+E(\Psi)-E(\Psi')\str } \gg 1,
\eeq
because the energy spacings are of order $W/d_{h}$ and $\langle \Psi\eta \str H_1\str \Psi'\eta'\rangle \sim g/\sqrt{d_{h}} $. Hence the ratio in~(\ref{eq: ratio}) is huge since $d_{h}$ grows exponentially in $n$. 

The outcome of the above calculation should not come as a surprise: it merely expresses that an ergodic bubble can act as a bath for a small system (here grain $y$) that is coupled to it. 
Upon repeating the same calculation a few times, one easily convinces oneself that states with the bubble in different positions hybridize with each other. {This in turn implies that they should appear with comparable amplitudes in typical eigenstates.}

\subsection{Spatial range of direct hybridizations} \label{sec:range}
In the above derivation, we focused on hybridizations that result in the translation of a bubble by one grain. One might worry that this is too negligible a translation if the bubble is very large, $n\gg 1$. However, here we show that direct hybridizations can take place at distances which are a finite fraction of the bubble length.  

As already pointed out, in the above derivation, we were careful to pick states $\eta,\eta'$ whose energy difference was small enough so that $\str\langle \eta, O_c \eta' \rangle \str \sim 1$. This is, however, not crucial, {and if $r'(y)=i,h$, then it cannot be assured anyhow}. The matrix element $\str\left< \eta\right| O_c\left| \eta' \right> \str$ will typically decay exponentially in the energy difference $E(\eta)-E(\eta')$. Hence, it can be as small as $\e^{-l_{0}\ep_m}$, but obviously this number decreases with $\ell_0$ but not with  $n$, so it cannot compete with the latter, if the bubble is sufficiently large. 
To determine at what distance direct hybridizations are possible, we proceed as follows.
Instead of making the transition $\eta \rightarrow \eta'$ at grain $y$, we now make a transition $\underline{\eta} \rightarrow \underline{\eta'}$ in a stretch of $\ell$ grains starting at $y$. 
By the structure of localized states, we know that 
\beq
\str \left< \underline{\eta}\right| O_c\left| \underline{\eta'} \right> \str \sim  (g/\ep_m)^{\ell \ell_0}.
\eeq
The transition is possible as long as this small number is larger than $\sqrt{1/d_{h}}$, so that we find 
\beq \label{eq: estimate range}
\ell \sim  \frac{s}{2 \ln (\ep_m/g)}  n.
\eeq
This shows  that the bubble hybridizes directly with a bubble configuration translated by a finite fraction of its size. However, this fraction becomes parametrically small as the coupling becomes weak $g/\ep_m\to 0$.

% SECTION
\section{Discussion of potential caveats}\label{sec: caveats}

We now discuss in more detail the caveats introduced in Section \ref{subsec: caveats}.

\subsection{Can bubbles {spread and permanently localize?}}

We have not yet added all coupling terms from $H_1$. Indeed, not only can the bubble move through the cold background, it can also spread its energy. Entropically, this is of course much more likely in real-time dynamics.
In particular, one sees that starting from a bubble configuration, the most likely thing to happen dynamically is that the bubble spreads until its energy density is intermediate or just below the putative mobility edge. At that point we cannot expect it to spread further as the involved states are now localized. The question arises as to whether these further couplings may induce a localization of  bubbles, despite the above construction of an apparently resonant, delocalized network of bubble configurations.
{We address this issue in two steps.   First, in Sec. \ref{sec: persistence}, we argue that a scenario in which  bubbles disappear dynamically is inconsistent. This is a conceptual  point.  Then, in Sec. \ref{sec: robustness}, we examine the hybridization argument on a formal level and we exclude that bubbles get localized by quantum dynamical effects. In Sec. \ref{sec: retardation}, we discuss how the fact that bubbles tend to spread slows  down their motion.}

\subsubsection{Persistence of bubbles} \label{sec: persistence}

Let us write $\langle \cdot \rangle_{\epsilon}$ for the expectation value in a microcanonical ensemble at energy density $\epsilon$ ({containing a large but subextensive number of eigenstates}). Let $n$ be the minimal length (in units of grains) of a well-ergodic bubble, as considered above. The thermodynamic probability of having such a bubble around position $x$ is given by
\beq \label{def: bx}
p_x = \langle  B_x \rangle_{\epsilon}, \qquad    B_{x} \equiv \chi(H_{[x-n/2, x+n/2 ]} \geq \ep_{c2} n\ell_0  ), 
\eeq
i.e.\ $B_x$ is the indicator of a local fluctuation at sufficiently high energy density.  Of course, $p_x$ becomes very small as $\ep_{c2}-\ep$, or $n$, or both are taken large, but it remains finite and independent of the total volume. By definition of the microcanonical ensemble,  
\beq  \label{eq: px as eigenstates}
p_x = \sum_{\Psi}  \frac{1}{\caN}   \langle \Psi \str B_x  \str \Psi \rangle,
\eeq
 and this  means that {it is not possible that no eigenstate has any weight on bubble configurations}.  One possibility consistent with~(\ref{eq: px as eigenstates}) is that
 typical eigenstates have an appreciable weight, $p_x$, on bubble configurations at $x$. Another (extreme) possibility is that a fraction $p_x$ of eigenstates have weight nearly $1$ on bubble configurations, while the others have none.   Either way, this rules out the scenario that bubbles could be completely absent in the eigenstates. 
 
 One can also formulate {the persistence of bubble configurations} in a dynamical way. This is more appealing if one thinks of our arguments as pertaining also to the evolution of well-chosen initial states, that contain a definite bubble. 
{Let us show this in a special and simple case. 
Consider a model where particle density is a conserved quantity and there is a putative mobility edge as a function of density, even at infinite temperature. Two obvious examples are the models studied in Refs.~\cite{muellerros, SchiulazMueller}.   In such a case, a relevant ensemble is one which constrains the particle density $\frac{N}{L} \in [\rho-\delta, \rho+\delta]$ to be close to $\rho$, without any constraint on energy.  The advantage of this case is that the projector onto the equilibrium ensemble corresponding to this constraint, can be decomposed into a basis of initial states $\str s \rangle $ which are products over grains with a definite particle number on each grain,  such that the total density is indeed in $[\rho-\delta, \rho+\delta]$. 
 Now, this ensemble $\sum_s \str s \rangle \langle s \str$ is exactly invariant under the dynamics and hence we obtain that the expectation value for seeing a bubble at position $x$,
 \beq
 \sum_{s}    \langle s \str U^*_t   B_x U_t \str s \rangle      =    p_x, \qquad \text{for any $t$},
 \eeq
 is invariant under  time evolution, and remains finite at all times.
Here $B_x$ and $p_x$ are again the indicator of a bubble around $x$ and the thermodynamic probability of a bubble (defined analogously to~(\ref{def: bx})), and $U_t$ is the Hamiltonian time-evolution.

\subsubsection{Robustness of hybridization}  \label{sec: robustness}
In the notation of section \ref{sec: formal bubbles}, the relevant type of transitions are the following:
\beq
\left.  \begin{array}{ccc} ccc{hhhh}ccc  & \leftrightarrow   &  cccc{hhhh}cc   \\
\updownarrow  &    &  \updownarrow   \\
 \ldots \leftrightarrow \ldots ccii{iiii}iicc  \qquad    &    &   \qquad cccii{iii}iic   \ldots \leftrightarrow \ldots ,
\end{array}\right.    \label{eq: bag}
\eeq
where the states on the lower line represent a multitude of mesostates {in which the bubble has partially spread its high energy density}. Let us assume that those states do not communicate with each other. {This simplifying assumption favors maximally the possibility that the coupling to such states could localize the bubble and thus could invalidate our preliminary conclusion regarding delocalization. }
We now consider the two subspaces, each of dimension $d_{h}$, that correspond to the mesostates on the upper line, the eigenstates of which are hybridized by the perturbation $H_1$. Let us refer to them as left and right subspaces. We then couple each of them to a space of {spread bubbles}, having a dimension $d_{h}' \gg d_{h}$, and ask whether the perturbation $H_1$ is still able to induce hybridization between left and right subspaces. {Concretely, the subspace $\C^{d_{h}}$ is now embedded in the space
$\C^{d_{h}}\oplus \C^{d_{h}'}$ of dimension $D_{h}\equiv d_{h}'+d_{h}$, and {the inter-grain coupling operator} $O_h$ becomes $O_h \oplus 0$. } {We focus on the transitions between the ergodic states $\Psi, \Psi'$ (notation as above), and just consider the operator $O_h$ which acts on the hot bubble.}
 Let us assume that after diagonalizing within the larger spaces of dimension $D_{h}$,  the eigenstates $\widetilde\Psi,\widetilde\Psi' $  are completely ergodic within those spaces and well captured by random matrix theory. ({In practice, this defines the relevant space to which the bubble subspace should be extended, and its dimension $D_{h}$.})  
 We now have to discuss how the ratio
 \beq \label{eq: ratio2}
\frac{\str \langle \widetilde\Psi \str O_h\str \widetilde\Psi'\rangle\str}{   \str \Delta E+E(\widetilde\Psi)-E(\widetilde\Psi')\str }
\eeq
differs from the original ratio
 \beq \label{eq: ratio1}
\frac{\str \langle \Psi \str O_h\str \Psi'\rangle\str}{   \str \Delta E+E(\Psi)-E(\Psi')\str }  \sim  \frac{\sqrt{d_{h}}}{W}
\eeq
with given $\str \Delta E\str \leq W$.
We find a suppression of the numerator  because now {
\beq
\str \langle \widetilde\Psi\str O_h \str \widetilde\Psi' \rangle\str \sim \frac{\sqrt{d_{h}}}{D_{h}}.
\eeq
}
Indeed, the simplest way to derive this is by remarking that 
\beq
\sum_{\widetilde\Psi, \widetilde \Psi'}   \str \langle \widetilde\Psi\str O_h \str \widetilde\Psi' \rangle\str^2=    {\rm Tr} (O^{\dagger}_h O_h)  \sim d_{h},
\eeq
as $O_h$ acts only in the original subspace (with dimension $d_{h}$) and it is zero on the attached space with dimension $d_h'$.
On the other hand, the energy spacing $\str \Delta E+E(\Psi)-E(\Psi')\str$ can now be made as small as $\widetilde W/D_{h}$, where $\widetilde W$ is the  width in energy of all states that significantly couple to the original bubble states. It follows that the ratio~(\ref{eq: ratio2}), and hence~(\ref{eq: ratio}), is  reduced by a factor $W/\widetilde W$. If this effect rendered the ratio~(\ref{eq: ratio2}) smaller than $1$, the eigenstates would likely not hybridize across the subspaces, i.e.,\ we would find \emph{localization induced by coupling to further degrees of freedom}. However, the maximal conceivable value of $\widetilde W$  is of order $ \epsilon_m \ell_h$, with $\ell_h$ the length of the region to which the energy spreads. Energy conservation and localization below $\epsilon_c$ lead to the upper bound $\ell_h (\ep_c-\ep) \leq n(\ep_m-\ep)$ (recall that $\ep <\ep_{c}$ is the typical energy density in our system). This yields {$\widetilde W/W \lesssim C$ with $C$ independent of $n$}. This is insufficient for localization, since the ratio~(\ref{eq: ratio}) is exponentially large in $n$.
 Thus, the hybridization of  states with large enough bubbles survives, despite their spreading to entropically more favorable states.
 This contrasts with {single} particle problems where the coupling to extra degrees of freedom was found to induce localization under certain circumstances~\cite{XieMueller,huselocbybath}. In those cases, there is no exponentially large factor that offsets the effect of an increased bandwidth $\widetilde W$, which renders coupling-induced localization possible.

  \subsection{Dynamic retardation}\label{sec: retardation}
Even though the inclusion of the states on the lower line of~(\ref{eq: bag}) cannot prevent hybridization, it does of course increase the \emph{timescale} necessary for transitions between the two bubble positions. The transition rates can be estimated from a simple Fermi Golden Rule calculation as
\bea \label{eq: fermi golden rule}
\tau_{\rm bef}^{-1} \sim \frac{\str \langle \Psi\eta \str H_1\str \Psi'\eta'\rangle\str^2}{   \str \Delta E+E(\Psi)-E(\Psi')\str }, \\
\tau_{\rm aft}^{-1} \sim \frac{\str \langle \widetilde\Psi\eta \str H_1\str \widetilde\Psi'\eta'\rangle\str^2}{   \str \Delta E+E(\widetilde\Psi')-E(\widetilde\Psi)\str },
\eea
before and after including the extra states, respectively.  The first rate is of order $g^2/W$, while the second is of order $(d_{h}/ D_{h})g^2/\widetilde W $. 
Hence, by adding the new states, we have increased the timescale by order $ D_{h}/d_{h}$ (keeping only terms exponential in $n$). This is very intuitive: Transitions are now only possible from a fraction $d_{h}/D_{h}$ of all states, and accordingly it takes longer until a transition will be attempted. Alternatively, one can view this as follows:  For a large bubble close to criticality (with structure $cciiiiiiiiicc$)  the 'active' configurations of the type $cccchhhhcccc$ manifest themselves as large deviations, which occur with exponential rarity (in the bubble size). Yet, as shown above, they do lead to {percolating} hybridization of eigenstates, and hence to delocalization.

\subsection{Harmlessness of rare, \\ strongly disordered regions in 1 dimension}
So far we have tacitly assumed that  the existence of a global thermal phase at some energy density implies that any large enough finite region  has ergodic states at that energy density.
However, due to rare fluctuations of the disorder, it can happen that rare, large regions still have all their states localized as long as they are disconnected from the rest of the system. 
One may think that especially in $d=1$ such regions could block global transport and thus localize the system at low temperature.
Even though this effect further increases the time scale necessary for thermalization, we now argue that it does not prevent it. 

As a preliminary, we consider a fully localized system of length $L_0$ in contact with an ergodic system of length $L_1$. By an analogous argument as used above for resonant delocalization, we see that for $L_0/L_1$ smaller than some number [depending on the localization length and the entropy density of the ergodic system, see e.g. Eq.~(\ref{eq: estimate range})] the coupled system will be ergodic: all formerly localized states can hybridize with each other.

{Now to the main argument. Let us  consider ergodic bubbles of some large size $\ell$. Now consider rare  regions of exceptional disorder that could block such bubbles, since an adjacent bubble of size $\ell$ would not be able to heat up this region to ergodic states (of type $h$).
Let  $\delta = \delta (\ell)$ be the typical distance between such rare regions.  By the above preliminary remark, such blocking regions have a length $\sim \ell$. However,  being rare regions (large deviations), the typical distance between them is much larger, $\delta (\ell) \sim \ed^{c\ell}$ for some $c>0$.   }
%One might think at first sight that the density of such blocking regions should be compared to the density of bubbles, but that is a fallacy. The point is that once there is a 
Now consider a bubble between two blocking regions. It renders the \emph{whole region} between them ergodic.
%, even though its thermalization time is very large. 
Hence, the blocking regions are in fact next to an ergodic bath of length $\delta (\ell)$, which is exponentially large in $\ell$. 
Accordingly, its level spacing is double-exponentially small in $\ell$. Thus, tunneling under the barrier of thickness $\sim \ell$, which is only exponentially small in $\ell$, will easily hybridize the ergodic regions on either side and ensure transport.

\subsection{Bubbles and weak localization in low dimensions}

As is well known, in dimension $d\leq 2$, non interacting particles are weakly localized by any non-vanishing disorder strength. Since bubbles resemble a particle-like excitation, we should discuss whether they undergo a similar weak localization in low dimensions. 

Indeed, at zero temperature, the answer is expected to be positive.
However, since we consider a finite energy density, it would be incorrect to picture the bubble as moving in a low-dimensional fixed disorder potential. As the bubble moves, it can excite or relax degrees of freedom. Thus, the Hilbert space locally resembles a tree, rather than a low-dimensional lattice (the number of relevant configurations that can be reached as the bubble moves grows exponentially, rather than polynomially with the traveled distance), and thus weak-localization effects should become irrelevant.  

More concretely, let $\ell$ be the localization length of the bubble motion at zero temperature, i.e., in the ground state (fixed environment). Obviously, $\ell$ increases with the bubble size since larger bubbles have more internal states (this is analogous to the increase of the single particle localization length with the number of channels in $d=1$); for large bubbles $\ell$ will be due to weak-localization effects.  Now, consider finite energy density, and $d=1$ for simplicity. Let $s>0$ be the entropy density in the cold background. Then, the condition $s \ell \gg 1$ is sufficient to ensure that inelastic scatterings of the bubble occur before the weak localization manifests itself, rendering them irrelevant for large enough bubbles. Note, however, that this condition places an additional lower bound on the size of mobile bubbles in low dimensions. 

\subsection{Restrictions on the type of rare events that may lead to delocalization}
The reader may wonder whether certain rare events similar to the bubbles discussed here would not rule out the possibility of genuine many-body localization altogether, or (erroneously) imply the absence of mobility edges in single-particle cases as well. However, in two appendixes we explain that this is not so.

In Appendix~\ref{sec: quenched spots}, we contrast the bubbles discussed here (namely disorder-independent high-energy fluctuations) with rare spots of low disorder in an otherwise fully localized system. We show why the latter are benign and do not hamper Imbrie's approach~\cite{Imbrie} to demonstrate localization, whereas high energy fluctuations do destroy localization, if the system is ergodic at high temperature.
In Appendix~\ref{sec: single particle}, we explain why our considerations apply only to genuine many-body systems, and how the bubble construction fails when applied to a single-particle system.

% SECTION
\section{Numerical results}\label{sec: numerics}

\begin{widetext}

\begin{figure}[t!]
\includegraphics[scale=0.45]{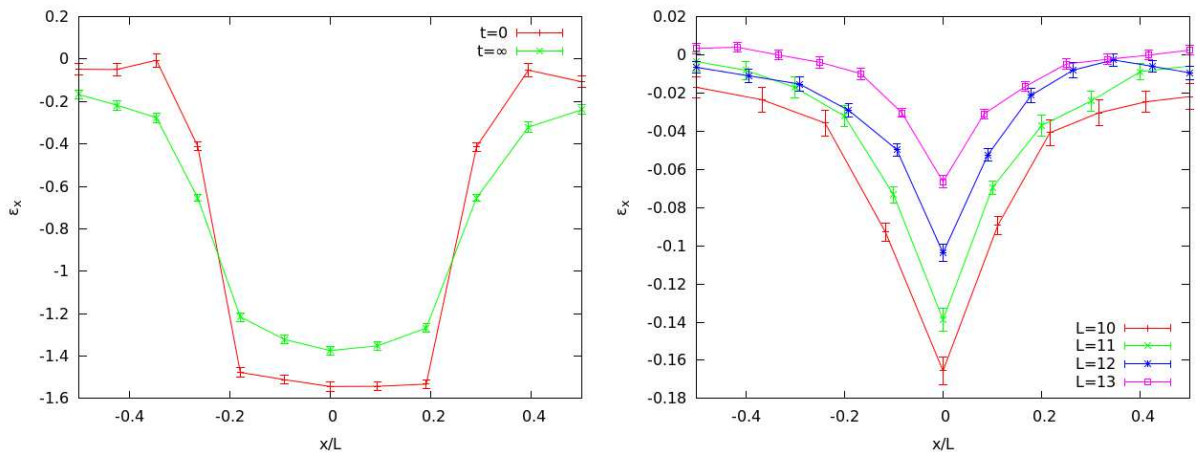}
\caption{{\em Left:} Disorder averaged energy per link $\varepsilon_{i}$ at $t=0$ (red) and averaged over time (green) for $L=12$. Initially a cold region of length $L_c=L/2$ is prepared. {The disorder strength is $\delta J=3J$. 
%The cold region is shown in the center of the plot for visual purposes. 
%The site index $i$ has been rescaled to the coordinate $x\equiv\frac{1}{L-1}\left(i-1\right)$. 
%The energy profile does not relax fully, even though the global energy density lies above the putative mobility edge. 
%This implies that, for small system sizes, a "hot" region is unable to act as a bath for the rest of the system. 
{\em Right:} Same protocol, but for $\delta J=J$ and very short cold intervals ($L_c=2$), at various $L$. }
The memory effects diminish with increasing $L$, but the hot region fails to thermalize the system well, even at the largest sizes. Results were averaged over 5000 disorder realizations.}
\label{Fig::protocolsAB}
\end{figure}

\end{widetext}

Our theoretical arguments contradict recent numerical data in favor of mobility edges~\cite{Pollman,Alet,Laumann}. The inconsistency is, however, only apparent. Indeed, we find that numerically accessible system sizes are not sufficiently large to host bubbles that are ergodic enough to be mobile. Therefore, delocalization by bubbles could not have been seen in numerics up to now.
In other words, the numerical results do not contradict delocalization by rare bubbles, but rather confirm that available sizes are not large enough.

\begin{widetext}

\begin{figure*}[t!]
%\centering
\includegraphics[scale=0.45]{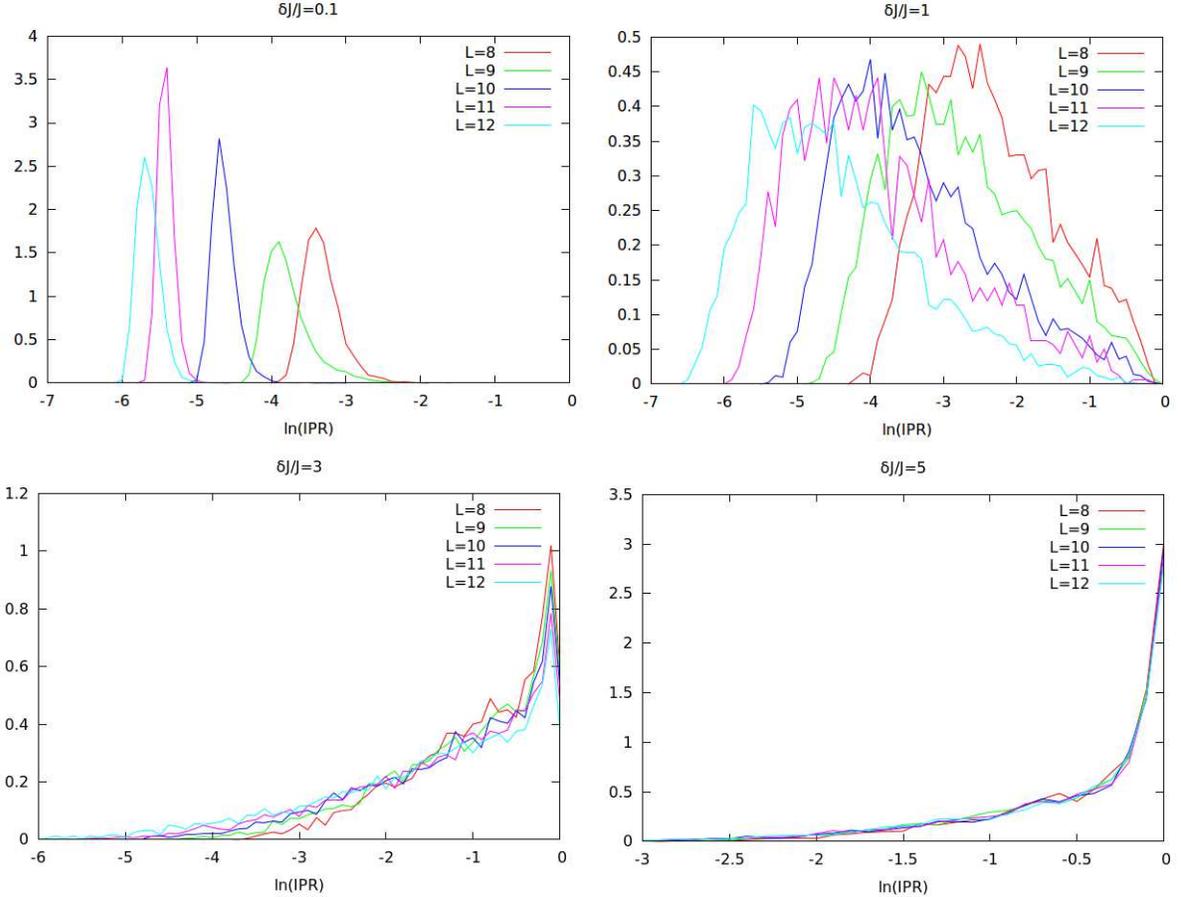}
%{IPR-eps-converted-to.pdf}
\caption{Distribution of $\ln({\rm IPR})$ associated with matrix elements of $\sigma_1^z$ evaluated on eigenstates randomly picked from the middle of the spectrum, for $\delta J/J=0.1,1,3,5$.  In the ergodic phase, the typical IPR is exponentially small in the size $L$. In the localized phase, the distribution is size independent. At $\delta J=J$ and the considered $L$,  the  typical IPR is exponentially small as in a truly ergodic phase, but the distribution is broad, and shows a tail towards localized values. $\delta J=3J$ is nearly critical: the 'localized' peak at ${\rm IPR} = O(1)$ slowly decreases with increasing $L$.}
\label{Fig::IPR}
\end{figure*}

\end{widetext}

We study the disordered Ising chain with next-to-nearest neighbor interaction considered in Ref.~\cite{Pollman},
\begin{eqnarray}
%H & = & -\sum_{i=1}^{L}\left(J+\delta J_{i}\right)\sigma_{i}^{z}\sigma_{i+1}^{z}+J_{2}\sum_{i=1}^{L}\sigma_{i}^{z}\sigma_{i+2}^{z}+\nonumber\\
% &  & h_{z}\sum_{i=1}^{L}\sigma_{i}^{z}+h_{x}\sum_{i=1}^{L}\sigma_{i}^{x},
H & = & -\sum_{i=1}^{L}\left[ \left(J+\delta J_{i}\right)\sigma_{i}^{z}\sigma_{i+1}^{z}+J_{2}\sigma_{i}^{z}\sigma_{i+2}^{z}+ h_{z}\sigma_{i}^{z}+h_{x}\sigma_{i}^{x}\right],\nonumber
\end{eqnarray}
where $\delta J_{i} \in \left[-\frac{\delta J}{2},\frac{\delta J}{2}\right]$ are independent random variables, and periodic boundary conditions are {taken}. We choose parameters
$J=1$, $J_2=0.3$ and $h_x=0.6$ as in Ref.~\cite{Pollman}, but add a finite $h_z=0.1$ to remove the Ising symmetry and the associated degeneracies.
The phase diagram in Ref.~\cite{Pollman} predicts a mobility edge in the thermodynamic limit at disorder strength $\delta J = 3$. To test our ideas,
we prepare the system at $\delta J = 3$ in a product state of the form $\left|\psi{(0)}\right\rangle _{L}=\left|\phi_{c}\right\rangle _{L_{c}}\otimes\left|\chi_{h}\right\rangle _{L-L_{c}}$, 
where $\left|\phi_{c}\right\rangle$ is the  ground state of an interval of $L_c$ sites, while $\left|\chi_{h}\right\rangle$ is an eigenstate of the complement close to the middle of the spectrum (a {hot} bubble). We choose $L-L_c$ as large as possible but such that the resulting global energy density is below the putative mobility edge. 
We then compute the time-evolving energy density on link $(i,i+1)$:
\begin{equation}\label{energy per site}
\varepsilon_{i}\left(t\right)\equiv-\left(J+\delta J_{i}\right)\left\langle \psi\left(t\right)\right|\sigma_{i}^{z}\sigma_{i+1}^{z}\left|\psi\left(t\right)\right\rangle.
\end{equation}
Our theory of mobile bubbles would predict that the $\varepsilon_{i}(t)$ profile becomes approximately flat as $t\to \infty$.
Via exact diagonalization, we evaluated its time average in finite system sizes,
 but almost no energy spreading from the initial state was observed [cf. Fig.~\ref{Fig::protocolsAB} (left)]. 
For tiny cold regions ($L_c=2$) and hot "bubbles" of almost the system size the global energy density is supercritical. Yet, still only a very small fraction of the bubble energy spreads to the cold region at $L=12$ (not shown),
while in the thermodynamic limit, the energy profile would obviously thermalize and become flat, as the system would be in its ergodic regime.
Therefore, these data show unambiguously that at our system size the hot region is {still} unable to act as a bath.

To document this further, we considered normalized states $\sigma_1^z \left|\alpha\right>$, with $\alpha$ an eigenstate, and calculated the inverse participation ratio (IPR) of its decomposition into eigenstates $\left|\beta\right>$ of the full system:
\begin{equation}
\textrm{IPR}_{\alpha}\equiv\sum_{\beta}\left|\left\langle \beta\right|\sigma_{1}^{z}\left|\alpha\right\rangle \right|^{4}.
\end{equation}
{The results are shown in Fig.~\ref{Fig::IPR}.}
At strong disorder, eigenstates are nearly eigenstates of $\sigma^z_i$ as well, and thus $\textrm{IPR}_{\alpha}\approx O(1)$, {with a distribution expected to become system-size independent for large $L$. As shown in the bottom-right panel of Fig.~\ref{Fig::IPR}, this is indeed the case for very strong disorder $\delta J=5J$, for which Ref.~\cite{Pollman} found the whole spectrum to be localized.} {Conversely}, deep in the delocalized phase, one expects eigenstate thermalization and behavior akin to random matrix theory,
$\left|\left\langle \beta\right|\sigma_{1}^{z}\left|\alpha\right\rangle \right|\propto \exp[-s L/2]$,
leading to a typical value $\textrm{IPR}_{\alpha}\sim \exp[-s L]$, with a  narrow distribution. {For very weak disorder, $\delta J=0.1J$, we found the exponent to be $s_{0.1}\approx 0.55$ which equals essentially the thermal entropy. The standard deviation of the distribution scales with system size in the same fashion, but is approximately 10 times smaller than the mean value. This is seen in the data of the left-top panel of Fig.~\ref{Fig::IPR}, where the relatively sharp peak in the distribution of $\log({\rm IPR})$ has an essentially $L$-independent width.

Let us now discuss close to critical disorder, $\delta J=3J$: the results shown in Fig.~\ref{Fig::IPR} confirm the absence of a truly ergodic phase {up to $L=12$}}. In fact, the distribution of IPR's at these parameters looks more characteristic of localization. Nevertheless, a slight, but clear tendency towards enhanced delocalization with increasing size is seen. This hints that in the thermodynamic limit the system will become ergodic, in agreement with the finite size extrapolation in Ref.~\cite{Pollman}.
To chart the lack of ergodicity at small sizes, we also look at $\delta J = 1$, where Ref.~\cite{Pollman} suggests that  most eigenstates are delocalized, even at $L=12$. Nevertheless, here, too, we find strong deviations from fully ergodic behavior, using the same two protocols as above. 
 Even in the extreme case of $L_c=2$ in Fig.~\ref{Fig::protocolsAB} (right), despite some energy transfer, the hot and cold regions are still clearly distinguishable after a long time evolution. To quantify this effect, we consider the time average of the energy imbalance between hot and cold regions,
$
\Delta\varepsilon\equiv
(L-3)^{-1}\sum_{i \notin \{c, c\pm 1 \}}(\varepsilon_{i}-\varepsilon_{c}),
$
where $c$ denotes the single link fully in the cold region. The imbalance decays exponentially with system size, $\Delta\varepsilon\sim\exp\left(-L/\xi\right)$ where $\xi$ increases with disorder strength. For ${\delta J/J}$ in the range $[1,1.5]$ we estimate $\xi\approx O(10)$ (see below), which sets a characteristic scale required to observe genuine ergodic behavior. This suggests strongly that at reachable sizes the {hot} bubble {is far from being} ergodic.

Also Fig.~\ref{Fig::IPR} illustrates that  systems with  $\delta J=J, L=12$ are far from the thermodynamic limit: the distribution of  $\ln\left(\textrm{IPR}_{\alpha}\right)$ is much wider (as compared to the mean) than in a clearly ergodic sample. To be more precise, the mean value still scales exponentially as $\textrm{IPR}_{\alpha}\sim \exp[-s_1 L]$, with $s_{1}\approx 0.5$, but the standard deviation is much larger than in the ergodic phase: indeed, the ratio between the standard deviation and  the typical value is of order $O(1)$. This means that there is a finite probability for a bubble to find itself in a position where the coupling to neighboring regions does not allow it to act as a bath. A considerable fraction of eigenstates will thus be localized. This illustrates that ETH, and hence our bubble arguments, cannot be applied for $\delta J=J, L=12$.

Finally, we studied the time average of the energy imbalance $\Delta\varepsilon$ as a function of system size and disorder strength $\delta J$. We considered  an initial state with a region of length $L_c=2$ in its ground state and the remaining system in an eigenstate near the middle of the spectrum.
 The results are shown in Fig.~\ref{Fig::Xi}, where each plot corresponds to a different disorder value. The results were averaged over 5000 disorder realizations. The data show that the imbalance
$\Delta\varepsilon$ decreases only slowly with system size. The dependence is consistent with an exponential decay. We have fitted the associated characteristic length $\xi_{\delta J}$, which grows with increasing $\delta J$.  These lengths $\xi_{\delta J}$ are of the same order as system sizes achievable in current numerical studies. Therefore, we conjecture that in order to be able to observe bubbles acting as good baths, one 
would need to study systems, which are larger by several $\xi_{\delta J}$'s. Perhaps the density matrix renormalization group (DMRG) techniques used recently in Refs.~\cite{DMRG1,DMRG2}, can shed new light on this.

In summary, the numerical analysis provided in this section clearly shows that the available system sizes are too small for ETH to be safely applied. Therefore, they are outside the range of applicability of our bubble argument. The numerical data present in the literature at the moment thus do not disprove our considerations.

\begin{figure*}
\includegraphics[scale=0.32]{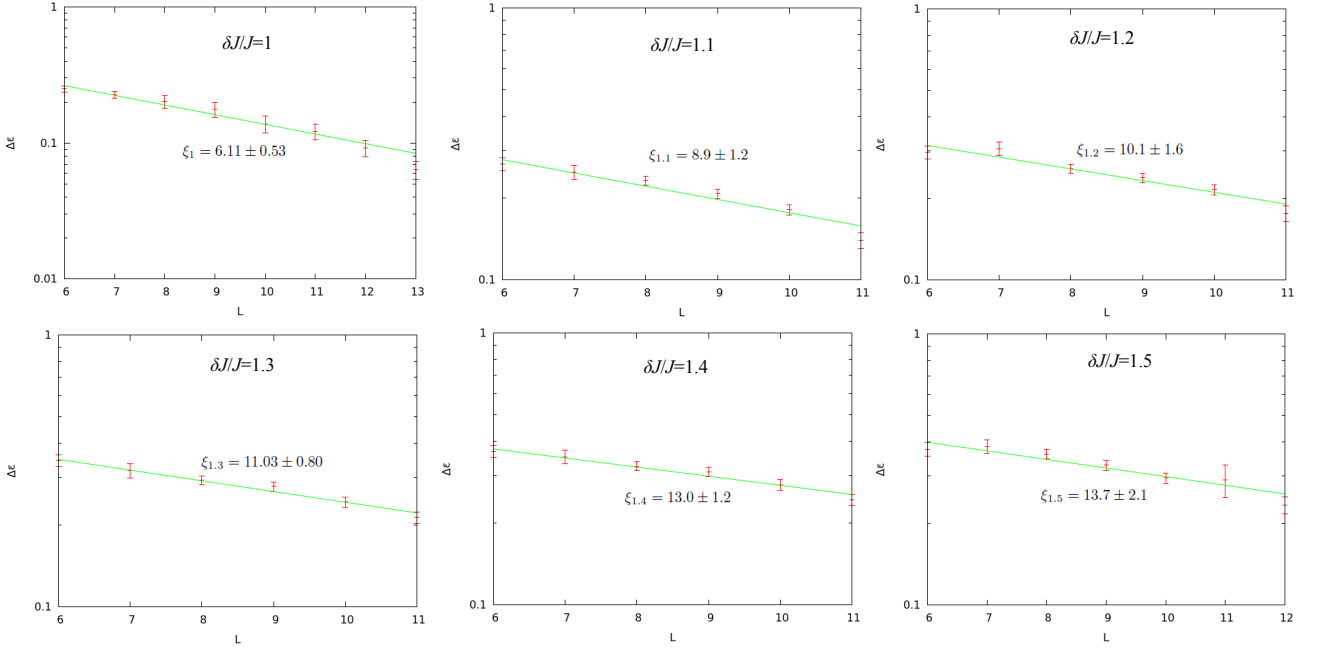}
\caption{Energy imbalance $\Delta\varepsilon$ as a function of system size, for disorder strengths $J\leq\delta J\leq 1.5J$. The curves are simple fits to exponentials. }
\label{Fig::Xi}
\end{figure*}

% SECTION
\section{Conclusion}\label{sec: conclusion}

We have argued that in the thermodynamic limit  many-body-localized and ergodic states cannot coexist, not even at very different energies. This has important consequences on the nature of the MBL transition. On a lattice, it implies that a transition is possible at best upon tuning the interaction strength, but not the temperature. In the continuum, genuine MBL is replaced by a strong crossover in the conductivity instead, which is notoriously hard to distinguish from a  genuine transition. Nevertheless, for practical purposes, on very long time scales, such badly conducting phases will behave as if they were genuinely many-body localized.

% SECTION
\begin{acknowledgments}
 
We thank D. Abanin, D. Basko, I. Gornyi, D. A. Huse, J. Imbrie, J. A. Kj\"all, A. Mirlin, R. Nandkishore, and A. Polyakov for useful discussions.
This research was supported in part by the National Science Foundation under Grant No. NSF PHY11-25915. 
M.M.\@ acknowledges the hospitality of the University of Basel, where part of this work was done. 
M.S.\@ acknowledges the hospitality of PSI Villigen.  W.D.R.\@ W.D.R. acknowledges the grant from the Deutsche Forschungsgesellschaft, DFG No. RO 4522/1-1. 
Both F.H.\@ and W.D.R.\@ acknowledge the support of the ANR grant JCJC, and thank the CNRS InPhyNiTi Grant (MaBoLo) for financial support.
\end{acknowledgments}

\appendix
\section{Assisted hopping model}\label{sec: assisted hopping}

Here we describe our numerical analysis for an assisted hopping model.  The main aim is to show that delocalization on a resonant subgraph remains robust to adding additional terms that connect that subgraph to localized states. We also show coexistence of localized and delocalized states, a failure of Mott's argument, which is, however, a particularity of the zero density limit of the considered model.

\emph{Description of the model.}
To reach the largest possible system sizes, we consider a Hamiltonian in $d=2$ with spin-orbit coupling, which gives rise to weak anti-localization and thus allows for a genuine delocalized phase. 
To the best of our knowledge, this is the smallest system where delocalization can be expected, {and is thus best suited for a numerical analysis}. Here, "smallest" means that 
the dimension of the Hilbert space grows at the slowest possible rate with growing linear size $L$. 

Let $H$ be the Hamiltonian of two indistinguishable hard-core bosons (with positions $q_{1,2}$) having a single spin-$\frac{1}{2}$ degree of freedom, $s$, {attached to them}. 
We consider points $q=(x,y)$ on the lattice $(\Z/L)^2$ and we impose periodic boundary conditions.  
The full Hamiltonian is
\beq
H = H_0 + h_1 H_1 + h_2 H_2,
\eeq
where $H_0$ is the uniformly distributed on-site potential
\beq
H_0 = \sum_{q} \epsilon_q a_q^+a_q, \quad -W \le \epsilon_q \le W.
\eeq
$H_1$ is the single-particle hopping Hamiltonian
\beq
H_1 = \sum_{q\sim q'} (a_q^+ a_{q'} + a_q a_{q'}^+),
\eeq
($q\sim q'$ denoting nearest neighbors)
and $H_2$ is the assisted hopping, including a spin-orbit interaction.
We describe $H_2$ by its matrix elements. 
Let 
\bea
\mathcal{S} & = & \left\{ q_{1}=(x_{1},y_{1}),\, q_{2}=(x_{2},y_{2})\,:q_{1}\ne q_{2},\right.\nonumber\\
 &  & \left.\max\{|x_{1}-x_{2}|,|y_{1}-y_{2}|\}\le1\right\} \label{S set}
\eea
be the set of pairs of {spatially neighboring} points. 
We then define $ \langle q_1',q_2',s' | H_2 | q_1,q_2,s\rangle $ to be
\beq
\mathbb I_{\mathcal S}(q_1',q_2') \mathbb I_{\mathcal S}(q_1,q_2) \, \langle q_1',q_2',s' | H_{\mathrm{SO}} | q_1,q_2,s \rangle,
\eeq
where the characteristic functions $\mathbb I_{\mathcal S}$ ensure that the initial and final pair configuration belong to $\cal S$.
Further, $H_{\mathrm{SO}} = H_{\mathrm{SO}}^1 + H_{\mathrm{SO}}^2$ with
%\begin{multline}
\bea
H_{\mathrm{SO}}^{1} & = & -i\left[\sigma^{(x)}T_{y_{1}}-\sigma^{(y)}T_{x_{1}}\right]\\
 & - & i\left[\frac{(\sigma^{(x)}-\sigma^{(y)})}{2}T_{x_{1}}T_{y_{1}}-\frac{(\sigma^{(x)}+\sigma^{(y)})}{2}T_{x_{1}}T_{y_{1}}^{\dagger}\right]+\text{H.c.}\nonumber
\eea
%\end{multline}
Here, $\sigma^{(x,y)}$ are Pauli matrices acting on the spin degrees of freedom, while 
the translation operators are defined by  $T_{x_1} \big|(x_1,y_1), (x_2,y_2),s\big\rangle = \big|(x_1+1,y_1), (x_2,y_2),s \big\rangle$ and similarly for $T_{y_1}$. $H_{\mathrm{SO}}^2$ is defined analogously.

The Hamiltonian $H_{\mathrm{SO}}^1$ is a lattice version of the Rashba Hamiltonian $\sigma^{(x)}p_{y_1} - \sigma^{(y)}p_{x_1}$.
We notice that restricting the definition of $H_{\mathrm{SO}}^1$ to the first term $-i\{ \sigma^{(x)} T_{y_1} - \sigma^{(y)} T_{x_1} \}$ would lead to a degeneracy due to the lattice structure. 
This would prevent $H$ from being a generic GSE Hamiltonian for any value of $h_2$.

\emph{Numerical results.}
In all the simulations, we take $L=9$ and $W=1$.
The analysis is divided into two parts:

(i) Delocalization via assisted hopping. First we take $h_1 = 0$ and $h_2 > 0$ (only assisted hopping). 
Since the majority of states (all configurations outside $\cal S$) are now trivially localized, we restrict ourselves to the subspace $\mathcal H_{\mathcal S}$ spanned by all the classical states in $\mathcal S$ (see~(\ref{S set})), each coming with spin up/down. 
We aim at finding $h_2$ such that $H_0 + h_2 H_2$ can be considered a ``typical" GSE matrix with truly delocalized eigenstates. 
For this, we evaluate numerically the parameter $r$ defined as
\beq
r = \Big\langle \frac{1}{\mathrm{dim}(\mathcal H_{\mathcal S}) - 2} \sum_{n=2}^{\mathrm{dim}(\mathcal H_{\mathcal S}) - 1} \frac{\min \{\Delta E_{n},\Delta E_{n-1}\}}{\max  \{\Delta E_{n},\Delta E_{n-1}\}} \Big\rangle\, ,
\eeq
where $\langle \cdot \rangle$ is the disorder average and $\Delta E_n\equiv E_{n+1}- E_n$, with $E_n$ being the ordered eigenenergies of the system. 
For the three classical ensembles, they take the values 
\beq
 r(\text{GOE})\simeq 0.53, \quad r(\text{GUE}) \simeq 0.60, \quad r(\text{GSE})\simeq 0.67.
\eeq
% Poisson: r=0.38
For $h_2=0.7$, we find $r=0.64 \pm 0.05$.
This value is significantly larger than $r(\text{GUE})$. {The discrepancy with $r(\text{GSE})$ presumably arises from the contributions from the more localized edges of the spectrum.}

To characterize (de)localization we use the logarithm of the inverse participation ratio,
\begin{equation}\label{IPR}
\mathrm{logIPR}(\psi) \equiv -\log_{10}  \Big(\sum_{\eta}|\langle \psi|\eta\rangle|^4\Big),
\end{equation} 
where the sum over $\eta$ runs over the classical particle configurations. 

%\MM{I don't understand the logics of the following sentences: the small logIPR probably come from the edges of the spectrum, but these edges should always host the same fraction of states, assuming that the DOS is size independent. I think there is a pollution of $r$ from those edges, independently of the size. On top of that there may be finite size effects in the sense of a stronger localization tendency at a given energy.}
Note that $\mathrm{dim}(\mathcal H_{\mathcal S}) = 648$, and thus $\text{logIPR}(\psi) \sim 2.5$ for a fully delocalized state $\psi$. From the point of view of the parameter $r$, $h_2=0.7$  is rather optimal: The spectrum is mostly delocalized, but the Hamiltonian is still genuinely GSE.
Indeed, when $h_2$ becomes significantly larger than $0.7$, the localized tails of the spectrum are further suppressed, but the value of $r$ starts bending down as an effect of approaching the integrable limit $h_2 \to \infty$. 

(ii) Robustness of delocalization against addition of single particle hopping.
 Let us now fix $h_2 = 0.7$, but vary $h_1 > 0$. 
We determine numerically the statistics for the logIPR's of the eigenstates $\psi$ of $H$.
The results are shown in Fig.~\ref{figure: IPR statistics}. 
The central message of that data is the following: Adding a finite $h_1$, which connects the resonant subspace $\cal S$ to its much larger localized complement, does not destroy the delocalization on the resonant subspace, as shown by the left and middle panels of Fig.~\ref{figure: IPR statistics}. 
In particular, for $h_1=0.07$ (middle) we  see delocalized states (inside the subspace $\mathcal H_{\mathcal S}$) coexisting with a majority of localized states. 
Obviously a relatively large $h_1$ leads to delocalization of almost all states, with logIPR's that start approaching the value $\log_{10}[\text{dim}(\mathcal H) = 6480] \sim 3.5$ of fully delocalized wave functions (cf. the right panel).
A comparison of histograms at the same values of $h_1$, but with $h_2=0$ (not shown) revealed that the histograms are significantly shifted to {larger logIPR in the presence of the delocalized channel of mobile pairs.} 

\begin{widetext}

\begin{figure}
\includegraphics[scale=0.45]{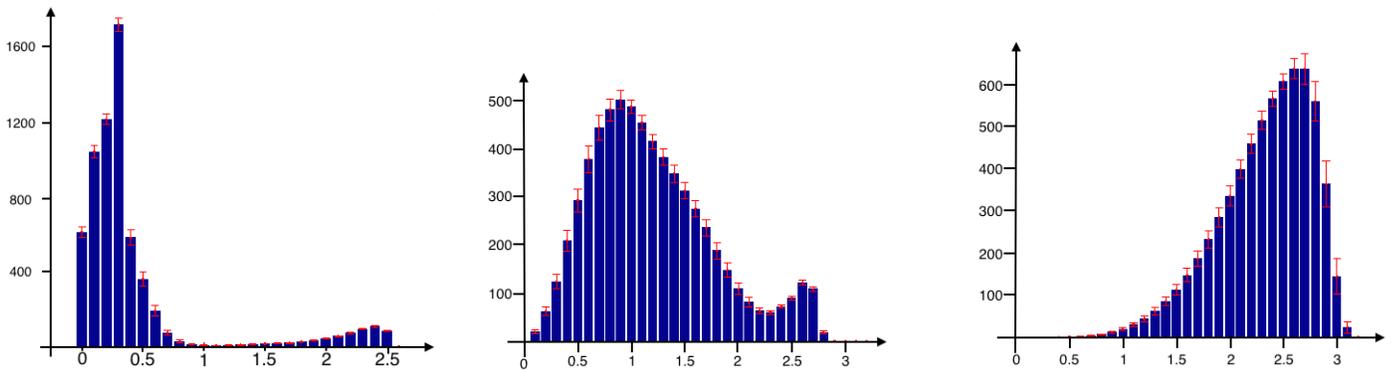}
\caption{
\label{figure: IPR statistics}
Statistics for the logIPR's of all eigenstates of $H$ for $h_2=0.7$ and different values of $h_1$.
From left to right: $h_1= 0.01$, $h_1=0.07$, $h_1=0.15$.  
Averages are taken over 500 realizations. 
}
\end{figure}

\end{widetext}

\section{Robustness of full MBL against the presence of localized ergodic spots}\label{sec: quenched spots}

It is instructive to compare the situation of high energy bubbles discussed in the main text with rare spots of low disorder in an otherwise fully localized system. In particular we explain here why the latter do not induce delocalization, even though a naive adaptation of some arguments we used in the main text might suggest so.

For simplicity we discuss systems in $d=1$. In the above reasoning we used the fact that, if an ergodic system of length $L_1$ is put in contact with a small localized system of length $L_0$ and if $L_0/L_1$ is smaller than a certain number, then localization does not persist: all formerly localized states hybridize with each other by mediation of the ergodic system they couple to. Now, one could naively think that this argument could be iterated {\em ad infinitum}: if one assumes that all states in the joint region of length $L_0+L_1$ are now fully ergodic, so that ETH holds, one could try to iterate the previous argument and conclude that yet a larger region would become ergodic, and so on. This would lead to the wrong conclusion that MBL systems simply cannot exist, because there is always a finite probability of ergodic inclusions of some finite size $L_1$.
However, a more careful analysis reveals where this reasoning is flawed, as we now show. 

Let us discuss a definite microscopic model to avoid ambiguities. Let $H= \sum_{x} H_x + JV_{x,x+1}$ be a Hamiltonian of a spin-$1/2$ chain,  where $H_x$ are on-site disorder terms with bandwidth $W$,  $V_{x,x+1}$ are hopping terms and  $J$ is a small coupling satisfying $J/W \ll 1$ (large disorder condition). 

Let $B:= [1,\ldots, L_1]$  be a local, finite-size "bath", i.e., a part of a chain which is ergodic if isolated from the rest, and $E:= [L_1,N]$ a localized region (the "environment").  This can arise, e.g., in a disorder realization where all $H_x$ for $1\leq x\leq L_1$ are equal up to deviations of order $J$, whereas such resonances are absent for $x>L_1$. 
Now split 
\begin{equation}
H=H_{B} + H_{B-E} +H_{E} 
\end{equation}
where the subscripts indicate the sets of spins on which they act (on $B$ only; both on $B$ and its complement $E$;  on $E$ only). 
%Concretely,
%$$
% H_{E}=   \sum_{x\in E} H_x + \sum_{ \{x,x+1\} \subset E} V_{x,x+1} 
%$$
Following Imbrie~\cite{Imbrie}, in the localized region $E$ we can find a unitary rotation that brings the Hamiltonian {$H_E$} into a canonical form of a sum of commuting terms, each being a product of "logical bit" operators. We now apply this rotation to the full Hamiltonian. Only the rotation of the terms which involve action on $B$ lead to terms that can flip the logical bits. However, their amplitude decays as $(J/W)^R$ where $R$ is the distance between the border of the ergodic region and the site at which the logical bit is centered. However, those flipping terms can induce significant hybridization between different values of the logical bits only if their amplitude is bigger than the level spacing in the region of diameter $L_1+R$, which scales as $2^{-L_1-R}$. Assuming sufficiently strong disorder $J/W< 1/2$, we define a buffer length $L_0$ by
\beq \label{eq: def buffer}
2^{-L_1-L_0} = (J/W)^{L_0}.
\eeq
The above reasoning implies that logical bits centered further than $L_0$ from the edge of the ergodic region are not hybridized through their coupling with the ergodic region $B$.
%\wdr{addition comes here}
To state this in a more formal fashion, we can split the chain into $\overline{B}:= [1,\ldots, L_1+L_0] $ and $\overline{E}=[L_1+L_0+1, N]$ and write the (rotated) Hamiltonian as 
\begin{equation}
H' =   H_{\overline{B} }+  H_{\overline{B} -\overline{E}}+ H_{\overline{E}}
\end{equation}
where the terms in $H_{\overline{B} -\overline{E}}$, acting on both $\overline{B}, \overline{E}$ are either commuting with $H_{\overline{E}}$ (terms diagonal in the logical bits), or have norm at most $(J/W)^{L_0}$  (terms originating from applying the rotation to $H_{B-E}$).  Using~(\ref{eq: def buffer}), this allows to conclude that the eigenstates of $H'$ are close to products of states in $\overline{B}$ and eigenstates of the logical bit operators in $\overline{E}$. 
 In other words, the delocalizing influence of the region $B$ is limited to a finite region of length at most $L_0$ around $B$, while many-body localization remains robust outside.

This shows that it is flawed to reason that the ergodic region $B$ can be extended to an ever enlarging neighborhood, as the bath would melt its neighboring degrees of freedom and incorporate them into the ergodic region. In fact, these enlarging regions gradually lose their full ergodicity. In particular, random matrix theory cannot  be applied naively when determining whether further degrees of freedom become delocalized as they are coupled to the already enlarged region.
 
In contrast, our bubble argument in the main text does not require such an iterative construction of a growing ergodic region. The basic difference is that a high-energy bubble can exist anywhere in space and is not tied to a rare quenched disorder configuration, which remains localized in space. We thus only need to argue that a hot bubble is able to displace itself in a finite series of elementary steps, thereby remaining ergodic inside. For those stages, we do not see reasons to doubt the applicability of random matrix-like behavior of matrix elements and level spacings. 

It is interesting to note that the above reasoning about the robustness of MBL cannot be generalized straightforwardly to $d>1$. The influence of ergodic spots in higher dimensional systems thus remains an interesting open question.

\section{Failure of our arguments for single particle mobility edges}\label{sec: single particle}

{It is  natural to ask why our argument for delocalization by bubbles does not apply to single-particle models, where we know that mobility edges do exist. In a nutshell, the argument does not apply because a single particle cannot borrow the energy from other particles to become hot and to delocalize at higher energy. Hybridizations between localized and delocalized states (as induced by some perturbation to the Hamiltonian) would necessarily require to go off-shell, which suppresses the coupling between localized states at large distance from each other. In contrast, the formation of a mobile bubble in a many-body system does not require the  violation of energy conservation.
Let us show that the natural analog of our argument fails. 
We have now a single particle Hamiltonian $H$ with localized eigenstates $\str \eta \rangle$ (essentially $\str \eta \rangle\simeq \str x\rangle$) for some $x$ and extended eigenstates $\str \Psi \rangle$, separated by a mobility edge at $E_c$. 
Upon switching  on a local perturbation $V$ (say, a change in the random potential) we evaluate the possible hybridizations.  Of course, $\str\eta\rangle, \str\eta'\rangle$ do not hybridize in first order. 
Also, $\str \eta \rangle$ does typically not hybridize with extended eigenstates $\Psi$. They are separated in energy and the matrix element $\langle \eta \str V \str \Psi \rangle \sim L^{-d/2}$ with $L$ the linear size of the system. Indeed, this is simply Mott's argument. 
(To be precise, one should check as well that the sum of all contributions extended contributions to the change in the state $\str \eta \rangle$ does not cause it to delocalize, but we skip this. The argument is very similar to what follows.)
However, the more interesting situation occurs in second order. Let us consider a typical term for the mixing of localized states $\eta,\eta'$ via extended states $\Psi$:
\beq \label{eq: second order}
\sum_{\Psi}   \frac{1}{E_\eta-E_\Psi}  \frac{1}{E_\eta-E_{\eta'}}    \langle \eta\str  V   \str \Psi \rangle \langle \Psi  V \str \eta'\rangle 
\eeq
A naive estimate gives that this sum is of order $L^{0}$, so potentially dangerous, but a more careful analysis refutes this: let us first change the sum over $\Psi$ by a sum over $\eta$. Then, we see that the resulting expression can become of order one only if $\str \eta\rangle $ and  $\str \eta'\rangle $ are close in space and their energy difference is comparable to the norm of $V$. This is obviously a case in which we \emph{do expect} $\str \eta\rangle $ and  $\str \eta'\rangle $ to hybridize. Hence, we exclude this case and we can freely add contributions of localized states to~(\ref{eq: second order}). Using functional calculus, we can recast the resulting expression as 
\beq \label{eq: second order again}
  \frac{1}{E_\eta-E_{\eta'}}    \langle \eta\str  V \frac{\chi( H)}{E_\eta- H}  V \str \eta'\rangle 
\eeq
with $\chi(E)$ a smooth cut-off function (we choose it to be a 'Schwarz function') that vanishes for $E<E_\eta +  (E_c-E_\eta)/2$ and becomes $1$ for $E\geq E_c$. It follows that
$$ \frac{\chi( H)}{E_\eta- H}=f(H) $$ for a smooth function $f$ of compact support.  Now we see that $f(H)$ is a local operator in the sense that
$$
\str \langle x \str f(H) \str y \rangle \str  \leq C(k) \str x-y \str^{-k}, \qquad \text{for any $k>0$}
$$ 
This is easily proven by Fourier transforming $f(H)=\int d t \e^{-i t H} \hat f (t)$, remarking that the Fourier transform of a Schwarz function is again Schwarz and using a 'Combes-Thomas' bound to get the locality  of $\e^{-i t H}$ on a spatial scale proportional to $t$.    For details on the functional analysis that was used in this argument, see for example \cite{reedsimon2}. 
The upshot is that~(\ref{eq: second order again}) is given by $  \frac{1}{E_\eta-E_{\eta'}}$ times an expression of order $\norm V\norm^2$ that decays rapidly in the distance between the localization centers of $\str\eta\rangle, \str\eta'\rangle$.  Hence, in the one-particle theory, the inclusion of extended states in perturbation theory does not destroy the localized states.


\begin{thebibliography}{99}
\bibitem{foot1} We believe that, in the thermodynamic limit,  the presence of finite conduction implies ergodicity, too. This follows if one disregards the exotic scenario of non-ergodic, but delocalized many-body systems, where transport is confined to a fractal support in real space. Indeed, in large enough systems, delocalized modes supported on a finite fraction of space will serve as a bath, which is expected to thermalize any finite set of degrees of freedom in finite times.
\bibitem{foot2} An analogous argument applies to disorder-free, but strongly interacting systems. There the putative transition can also be tuned by density and the obviously delocalized phase is at low density, see~\cite{bubbles}. The analogues of the hot bubbles discussed above are then bubbles of low density.
\bibitem{foot3} Note that, in contrast, the entanglement of the bubble region with its surroundings grows logarithmically without limits, reflecting that an increasing number of degrees of freedom affects the relative phases of configurations in the bubble region. However, this does not alter the fact that the relevant number of hybridizing configurations remains finite.
\end{thebibliography}
\end{document}